# Debt-Prone Bugs: Technical Debt in Software Maintenance

Jifeng Xuan, Yan Hu, *He Jiang
*School of Software, Dalian University of Technology*
*Dalian 116621, China*
*xuan@mail.dlut.edu.cn, huyan@dlut.edu.cn, hejiang@ieee.org*

**_Abstract_**

*Fixing bugs is an important phase in software development and maintenance. In practice, the process of bug fixing may conflict with the release schedule. Such confliction leads to a trade-off between software quality and release schedule, which is known as the technical debt metaphor. In this article, we propose the concept of debt-prone bugs to model the technical debt in software maintenance. We identify three types of debt-prone bugs, namely tag bugs, reopened bugs, and duplicate bugs. A case study on Mozilla is conducted to examine the impact of debt-prone bugs in software products. We investigate the correlation between debt-prone bugs and the product quality. For a product under development, we build prediction models based on historical products to predict the time cost of fixing bugs. The result shows that identifying debt-prone bugs can assist in monitoring and improving software quality.*

## 1. Introduction

Software bugs are common and unavoidable in development and maintenance. In modern software industry, many projects employ *bug tracking systems* to manage bugs. For example, Bugzilla (*bugzilla.org*) is a widely-used bug tracking system in open source projects, which has been deployed in over 1200 projects. Once a software product is under testing, bugs accumulate to the bug tracking system. However, the efforts of fixing bugs may conflict with the release schedule. On one hand, developers spend much cost (including labor and time) in fixing bugs to improve the quality; on the other hand, the schedule of releasing the product restricts the cost of bug fixing. Such conflict leads to a trade-off, i.e., only fixing a part of bugs to follow the release schedule or fixing all the bugs to ensure the quality by delaying the release schedule.

The trade-off between the release schedule and the software quality is known as technical debt. The technical debt, a metaphor, was first coined in 1992 to denote the trade-off between long-term code quality and short-term gain [2]. After the evolution in the past two decades, technical debt has been extended from code to various types, such as architecture debt, design debt, and maintenance debt [1]. Exploring technical debt becomes an effective approach to understanding the finite cost in development [7] and to measuring the software quality [4], [8].

In this article, we propose the concept of debt-prone bugs to model the technical debt of products. A debt-prone bug, which is the debt in maintenance, is produced by an incomplete or immature process of bug fixing and can add risks to software quality. We examine three types of debt-prone bugs, including tag bugs, reopened bugs, and duplicate bugs. To investigate debt-prone bugs of software products, we conduct a case study on the Mozilla project. For each type of debt-prone bugs, three attributes are extracted, namely the number of bugs, the frequency of debt-proneness, and the time cost of fixing bugs. We consider leveraging the debt-prone bugs to predict the average time of fixing bugs for a product. By analyzing the correlations between debt-prone bugs and products, we build prediction models based on the debt-prone bugs of historical products. Given a product under development, the predicted results can guide the decision of maintenance for the project manager or developers.

The primary contributions of this paper are as follows.

1. We propose the concept of debt-prone bugs, which extends the existing technical debt in software maintenance;

---

* *Corresponding author*

2. We identify three types of debt-prone bugs of software products, namely tag bugs, reopened bugs, and duplicate bugs;

3. We conduct a case study on the Mozilla project to investigate the correlation between debt-prone bugs and software quality. We further build prediction models to predict the average time cost of fixing bugs for a given product. For industry, debt-prone bugs provide a platform to monitor and understand the software quality.

The remainder of this paper is organized as follows. Section 2 presents the background of bugs in a bug tracking system. Section 3 describes how to identify three types of debt-prone bugs. Section 4 conducts a case study on the Mozilla project. Section 5 states the threats to validity. Section 6 lists the related work. Section 7 concludes and presents the future work.

## 2. Background

A bug tracking system (also known as an *issue tracking system* or a *bug repository*) is a database for collecting and managing bugs to support software maintenance. A bug is assigned with an ID and starts its lifecycle once a developer or tester submits it to the bug tracking system. The submitter of a new bug describes the problem in natural languages and reports the environment, which may be helpful to find the root cause. Then any developer, who is interested in this bug, can make comments to facilitate the process of bug fixing, e.g., adding details for reproducing the bug or creating a patch. For a bug, the description and the comments direct support the process of fixing this bug. According to the process of fixing bugs, developers label a bug with various statuses, including new, fixed, and invalid, etc (in this article, the term *fixing* is used in a board sense, including adding a patch, identifying an invalid bug, or closing a never-fixed bug, etc.). Figure 1 presents a bug in the Mozilla project (a large-scale open source project, *mozilla.org*). This bug aims to add implementation to the layers for plugins (for details, see *https://bugzilla.mozilla.org/show_bug.cgi?id=556487*). In Comment 21, a developer proposes a patch to fix the bug; in Comment 24, another developer makes comments on this patch.

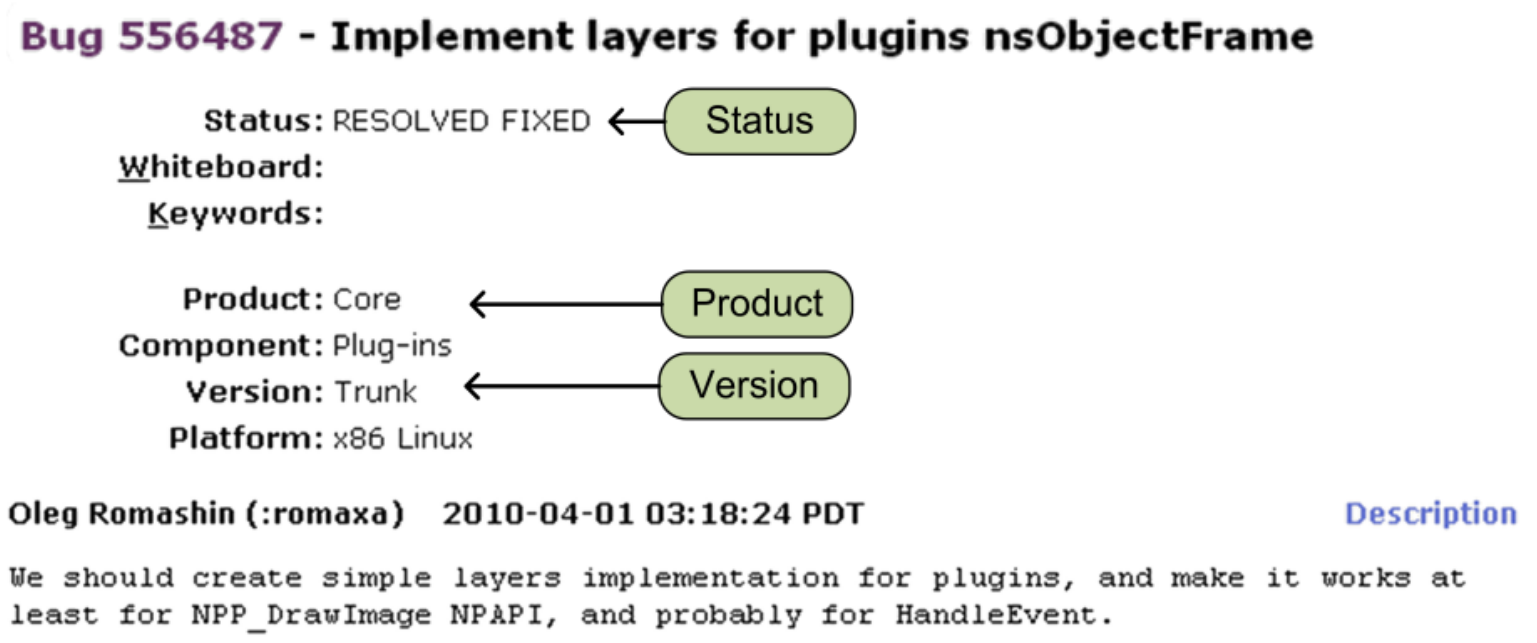

(a) A bug with its status, product, and version

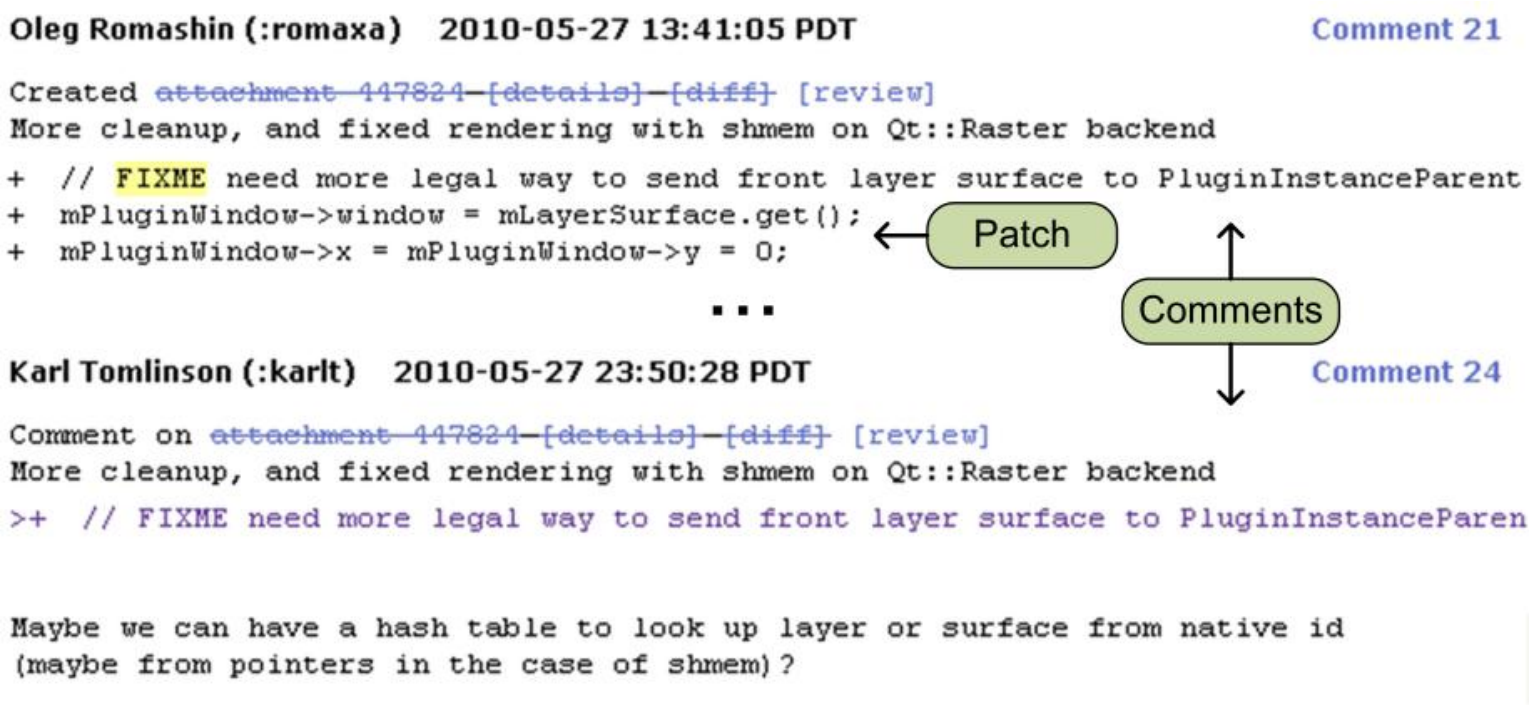

(b) Two comments on the bug

**Figure 1**. Snapshot of a bug in Mozilla and its two comments

Besides the status of a bug, the bug tracking system records the related product and its version. In a software project, a product is a sub-project, which can be either an individual software package (e.g., Product Firefox, a web browser in Mozilla) or a dependent model in the project (e.g., Product Core in Figure 1, a kernel model for other products). Many products have version numbers (e.g., Firefox 8.0) while several products may live in a long-term lifecycle and only have one version name trunk (i.e., a continuously existing product). To simplify the following statements, we use a product to denote a product with its version.

## 3. Identifying Debt-Prone Bugs

In this article, a debt-prone bug refers to a software bug, which is caused by an incomplete or immature process of bug fixing (e.g., inadequate efforts due to limited cost). Such bugs can be viewed as the debt in bug fixing, which can weaken the product quality and needs extra cost for further handling. For example, restricted by of the release time, developers have to fix bugs under pressure. This fact leads to incomplete bug fixing or even some new bugs, which are hidden in the release [17]. In our work, we present three types of debt-prone bugs, namely tag bugs, reopened bugs, and duplicate bugs.

### 3.1 Tag Bugs

In software development, developers usually use tags to annotate the code. A widely used tag is TODO. As its name suggests, a TODO tag indicates the task, which should be done in future work. Adding TODO tags to code is caused by various reasons, such as the lack of ideas for implementing a feature or the ending of a workday. Developers usually insert TODO tags to prompt the unfinished work. However, as the accumulation of TODO tags in code, some of the TODO tags may be forgotten, or even become bugs [9]. In our work, we label the bugs which are caused by such tags as debt-prone bugs. Besides the TODO tag, we check two other kinds of tags, i.e., FIXME and XXX. In Mozilla, a FIXME tag denotes the potential problematic code in the current implementation while a XXX tag denotes a bad style or structure in code. Taken the FIXME tag in Figure 1 as an example, the patch in Comment 21 adds a FIXME tag to denote the need of the correct implementation. Then in Comment 24, a new solution is proposed to handle such FIXME tag. Therefore, the patch in Comment 21 cannot completely fix the bug, but provides a new bug for further handling. In other words, a patch with a FIXME tag is a compromise between the maintenance cost and the potential problematic code. We denote a bug caused by the tag TODO, FIXME, or XXX as a tag bug, which is listed as the first type of debt-prone bugs.

### 3.2 Reopened Bugs

In bug tracking systems, a bug may be solved by developers, but reopened later. Such bugs are called reopened bugs. In a typical reopened bug, its status may change as follows, assigned to a new developer, fixed by the developer, reopened by another developer, and fixed by such second developer. In such a bug, unless the bug has been really fixed, it is hard to find out whether this bug is resolved. The causes for reopenings include the unclear description of bugs and the insufficient information for reproducing bugs. Existing study finds out that the time cost of reopened bug is longer than that of other bugs, especially bugs with multiple times of reopenings [5]. We consider reopened bugs, i.e., the bugs caused by the reopening, as the second type of debt-prone bugs. Note that in Mozilla, a reopened bug has only one bug ID, although it has been opened more than once. A reopened bug can refer to either the incompletely fixed bug (which causes the reopening) or the newly opened one. In this article, we focus on the latter one, i.e., the newly opened bugs, caused by former bugs.

### 3.3 Duplicate Bugs

A duplicate bug is a new bug, which has the same root cause of an existing bug (called a master bug) in the bug tracking system. Duplicate bugs can be divided into two categories. One contains the duplicate description of the same failure as a master bug; the other contains a different failure, which is originated from the same root cause as a master bug [3]. The first category is caused by immature search techniques while the second category suffers from the lack of knowledge about existing bugs [10]. Ideally, if a developer is familiar with all the existing bugs, the duplicate ones can be avoid. However in practice, developers cannot examine all the existing bugs to determine whether a new bug matches an existing one. Since duplicate bugs are caused by the inadequate examination, we consider duplicate bugs as the third type of debt-prone bugs. In the bug tracking system, a duplicate bug is assigned with a new bug ID, i.e., a new bug different from existing bugs. Moreover, in practice, multiple duplicate bugs may share one master bug. For example, if a bug B is the duplicate of a bug A, and a bug C is the duplicate of B, then A is the master of bugs B and C. It is helpful to explore the root cause of duplicate bugs by identifying the master bug [3].

Among the above three types of debt-prone bugs, a tag bug or a reopened bug shares the bug ID with an existing bug, which induces the debt to the products. Thus, tag bugs and reopened bugs will extend the time cost of bug fixing but not enlarge the number of bugs to products. In contrast, duplicate bugs will enlarge the number of bugs. Due to the rise of the number of bugs, the total time of fixing bugs is also enlarged by the duplicate bugs.

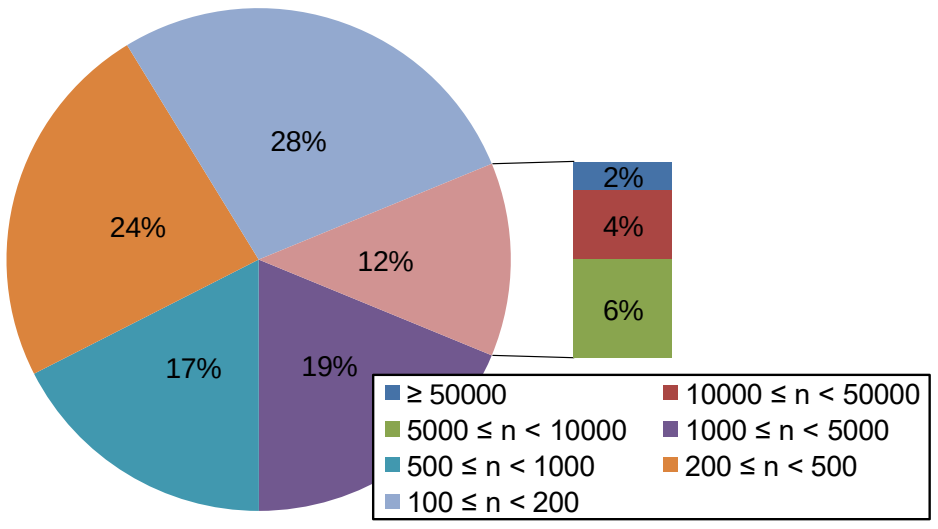

(a) Ratio of products according to the number of bugs in each product (denoted with n).

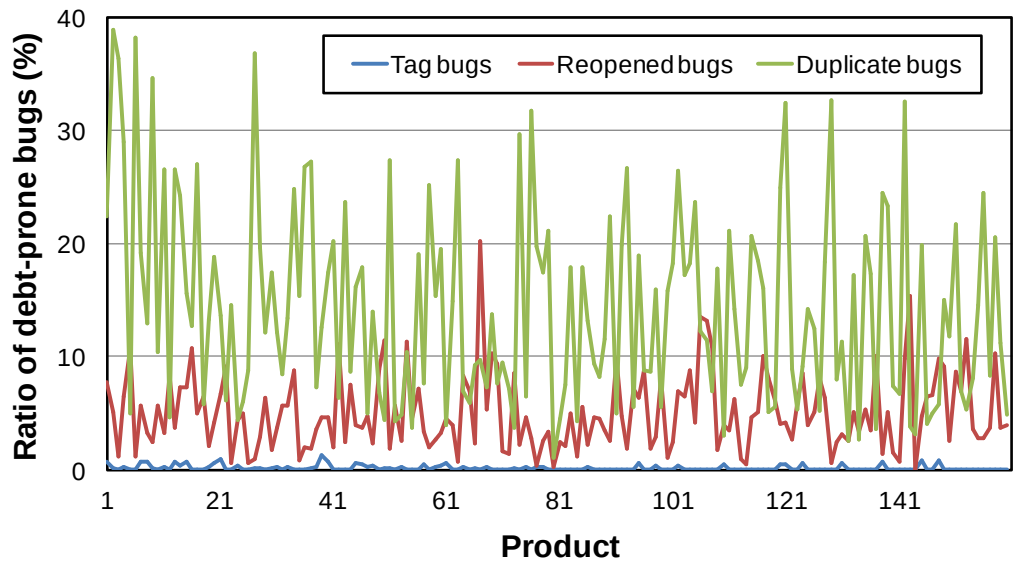

(b) Ratio of each type of debt-prone bugs in products

**Figure 2.** The 160 products and their debt-prone bugs in Mozilla

## 4. A Case Study on the Mozilla Project

Growing debt affects the quality of software products and slows down the development [15]. Measuring software bugs is a good method for monitoring the software quality [14]. In the following part of this article, we conduct a case study on the Mozilla project to investigate the impacts of debt-prone bugs on software quality. We collect all the bugs of Mozilla from 1998 to the end of 2010 and classify all these bugs according to their related products. We extract three types of debt-prone bugs of each product. For reopened bugs and duplicate bugs, we collect bugs according to the identifiers, which are provided by the bug tracking system; for tag bugs, we search the tags TODO, FIXME, and XXX in comments and manually validate whether these tags induce new bugs. In total, 599870 bugs and 567 products are collected. We remove the products with less than 100 bugs to avoid the interference of small-scale products. Then 160 products are left for the experiments. These 160 products consist of 596495 bugs, including 1895 tag bugs, 32482 reopened bugs, and 142643 duplicate bugs. In Figure 2, we present the statistics of products and debt-prone bugs in each product. As shown in Figure 2(a), 88% of products have more than 100 and less than 5000 bugs; as in Figure 2(b), duplicate bugs account for a larger percentage than the other two types of debt-prone bugs.

## 4.1 Framework

Bug fixing can be viewed as a similar activity of implementing requirements by adding new features to software products [12]. From the perspective of software maintenance, debt-prone bugs are factors of low quality, which are caused by incomplete software process; from the perspective of requirements engineering, debt-prone bugs restrict the improvement of existing products. In this article, we employ debt-prone bugs to understand the debt in development and to improve software products. We investigate attributes of debt-prone bugs in software products and predict the average time cost of bug fixing for a product under development.

First, we characterize the debt-prone bugs with three attributes, namely the number of bugs, the frequency of debt-proneness, and the time cost of fixing bugs. Given a product, we extract such three attributes for each type of debt-prone bugs to indicate the technical debt. The study on these attributes indicates that these attributes can impact the average time cost of fixing bugs in the product.

Second, machine learning algorithms (e.g., the linear regression) are employed to train prediction models based on attributes and average fixing time of historical products. For a product under development, we can collect its known debt-prone bugs for prediction models and then predict the potential average time of fixing all bugs. The predicted result can guide the project manager and developers to monitor the software quality and to make a decision on the future work. For example, if too much time is cost in fixing bugs, it is necessary to adopt better management and project schedule.

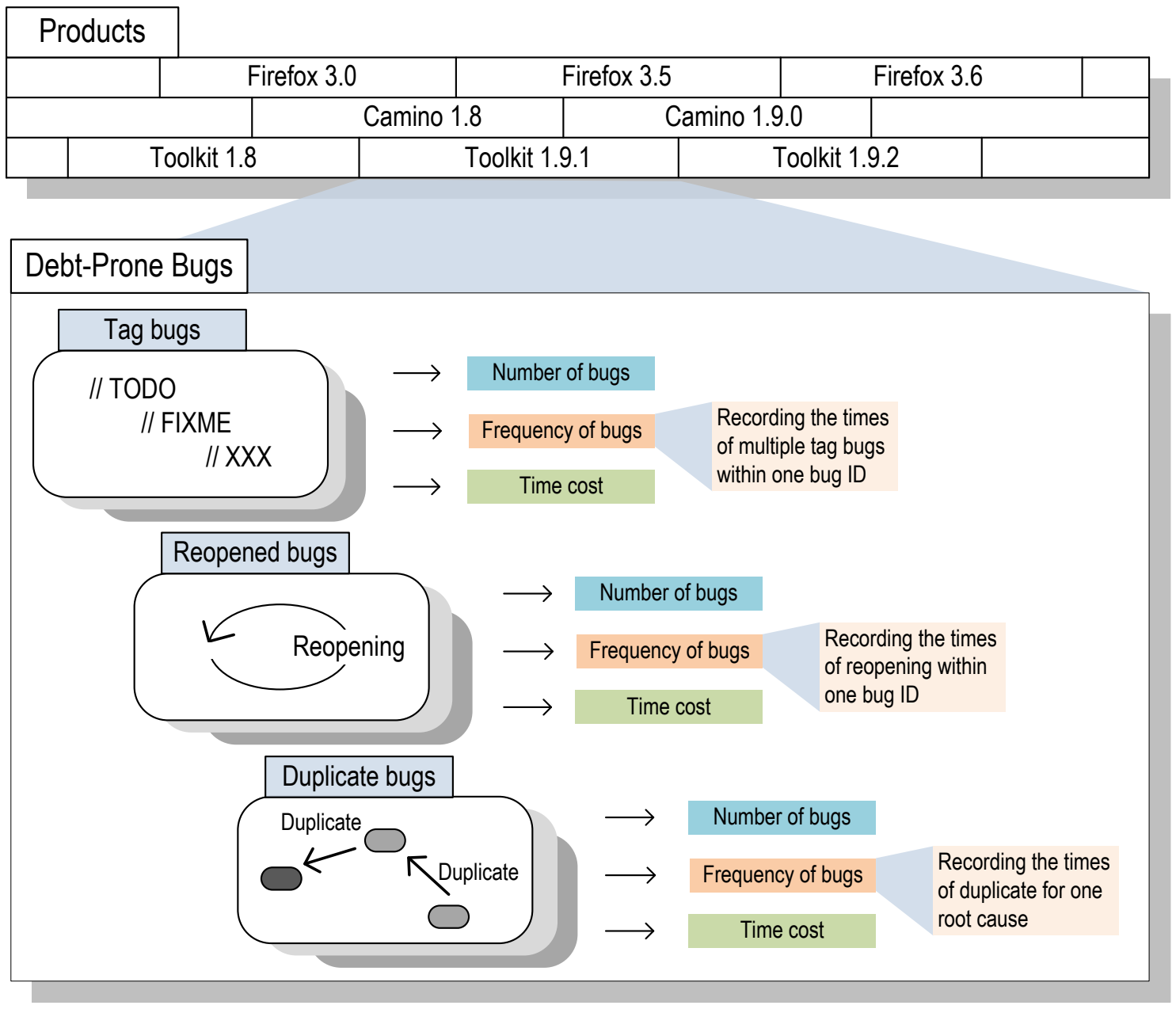

**Figure 3.** Extracting attributes for debt-prone bugs of software products

## 4.2 Attribute Calculation

In Figure 3, we present the framework of building prediction models with extracted attributes of debt-prone bugs. Among three attributes of each type of debt-prone bugs, the first attribute, i.e., the number of bugs, denotes the total number of bugs within this type. The second attribute, i.e., the frequency of debt-proneness, is calculated to denote the times of debts. For a tag bug, the frequency denotes how many tag bugs appear within one bug ID; for a reopened bug, the frequency denotes the number of reopenings in one bug ID; for a duplicate bug, the frequency is the number of duplicate bugs, which are due to the same root cause. We count the frequency for a product with the average frequency of all the bugs in one type. The third attribute is the time cost of fixing bugs, which is spent on fixing each type of bugs in this product. For one bug, we count days from the date when the bug is assigned to a developer to the date when the final change is done to the bug.

The attributes in our work is a kind of software metrics [16], [18], which is employed to measure the software quality. For each product, we extract the values of nine attributes and consider these attributes are helpful to model the quality of products in Mozilla.

## 4.3 Correlation Analysis

The time cost of bug fixing is an important indicator to measure the quality for software products [6]. We consider employing debt-prone bugs to predict the average time cost of fixing all the bugs in a product. We use the Pearson correlation coefficient [11] to investigate the relationship between the attributes of debt-prone bugs and the time cost in a product. Figure 4 presents the details of the correlation. As shown in Figure 4, the attributes, time cost of reopened bugs and time cost of duplicate bugs provide the strong correlation; time cost of tag bugs provides the modest correlation; frequency of tag bugs and frequency of reopened bugs provide the weak correlation. In other words, among all the attributes, the time cost of duplicate bugs can be viewed as the strongest correlated factor to the average time cost of fixing bugs in a product. A potential reason for this fact is that the ratio of duplicate bugs is larger than those of the other two types of debt-prone bugs.

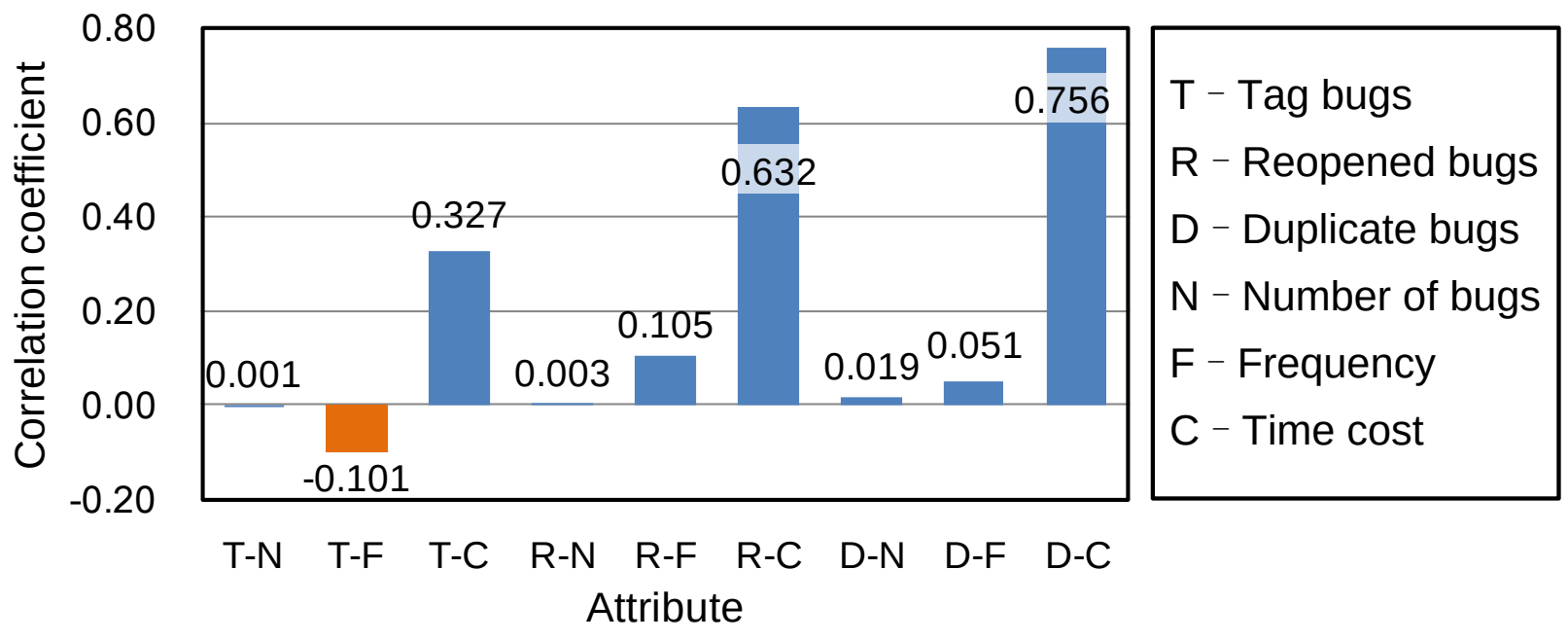

(b) Pearson correlation coefficient for each attribute

| Correlation | Negative | Positive |
|---|---|---|
| None | -0.1≤ r ≤0.1 | |
| Weak | -0.3≤ r <-0.1 | 0.1< r ≤0.3 |
| Modest | -0.5≤ r <-0.3 | 0.3< r ≤0.5 |
| Strong | -1.0≤ r <-0.5 | 0.5< r ≤1.0 |

(b) Correlation level for the value of coefficient

**Figure 4.** Correlation coefficient between attributes and average time cost of fixing bugs in products

## 4.4 Average Time Cost Prediction

We aim to use the attributes of debt-prone bugs to predict the average time cost of bug fixing for a given product. In detail, we first train prediction models with the existing products. Then, for a new product, e.g., a product under development or to be released, we collect all the known debt-prone bugs and use the above model to predict the time cost of fixing bugs. According to the predicted value, the project manager and developers can make a decision on the future work. In Table 1, we present the results of predicting the time cost of fixing bugs on 160 products. We examine the results of prediction models based on three machine learning algorithms, namely the multilayer perceptron (an artificial neural network), the model tree (a decision tree), and the linear regression (a linear optimization model). The evaluation of results is based on 10-fold cross-validation (an evaluation method) and the implementation is based on a machine learning tool, Weka [11]. Among the three prediction models, the linear regression has the best result, with the highest correlation coefficient (i.e., the most correlated values to the real ones) and the lowest root relative squared error (i.e., the least relative error). The predicted results by the model tree and the linear regression are acceptable, but have large relative error. To improve the accuracy of prediction, it is helpful to extract detailed attributes for characterizing the debt-prone bugs.

**Table 1.** Results of predicting average time cost of fixing bugs with three prediction models

| **Prediction model** | **Average time cost** | |
|---|---|---|
| | **Correlation coefficient** | **Root relative squared error** |
| Multilayer perceptron | 0.699 | 81.86 % |
| Model tree | 0.792 | 60.74 % |
| Linear regression | 0.804 | 59.10 % |

## 5. Threats to Validity

In this article, two aspects of threats should be further validated, namely the multiple-type debt-prone bugs and the prediction models based on debt-prone bugs.

First, we identify the debt-prone bugs to understand the technical debt of software products. We collect three types of debt-prone bugs to conduct the case study. However, several bugs may belong to multiple types. For example, a bug could be both a tag bug and a duplicate bug. In our study, the attributes of such multiple-type bugs are respectively counted and are overrated for each type of bugs. To avoid such overrating, we can either force each bug in only one type or propose a specified type to denote multiple-type bugs, which are shared in multiple types.

Second, we present the correlation between the debt-prone bugs and the quality of software products. We predict the average time cost of bug fixing for products with known debt-prone bugs. Since the predicted results cannot equals to the real values, we do not claim that the project manager or developers need to completely make predictions with our models. Based on the case study, we want to provide an approach to guide the manger or developers, i.e., how to monitor and understand the quality of products with debt-prone bugs. Moreover, our prediction models with debt-prone bugs can be transferred to predict other potential values of software quality.

## 6. Related Work

To our knowledge, no existing work has explored the technical debt by characterizing the bugs of software products. The related work of this paper can be divided into two categories, namely the technical debt in software maintenance and the software bug tracking system.

Guo et al. [4] have examined the technical debt in software maintenance with an exploratory case study on a real application. Zazworka et al. [15] describe how to identify the impacts of software quality with the technical debt. To further understanding the technical debt in software engineering, Brown et al. [1] summarize the state of art in managing technical debt in software systems and present the potential future work.

In bug tracking systems, Storey et al. [9], Wang et al. [10], and Guo et al. [5] have studied specified types of software bugs. Zimmermann et al. [14] examine the quality of bugs by questionnaires and design a predicted model to identify whether a bug is within high quality. Xuan et al. [13] investigate the developer prioritization in bug tracking systems and employ such prioritization to assist existing software tasks.

## 7. Conclusion and Future Work

Technical debt is important to understand the compromise between software cost and quality. In this article, we focus on the debt-prone bugs, which are the debt in maintaining software products. Based on a case study on Mozilla, we investigate three types of debt-prone bugs. Prediction models are built on historical debt-prone bugs to predict the average time of bug fixing for products.

On the basis of our findings in debt-prone bugs, we consider the technical debt as potential factors for indicating software quality. This motivates our future work on exploring other types of debt-prone bugs. Moreover, we plan to extract more effective attributes of debt-prone bugs and further investigate the usage of the debt in software maintenance. Another future work is to combine the debt-prone bugs with the project scheduling. We plan to explore how to employ the debt-prone bugs to improve the process of maintenance.

## 8. Acknowledgments

We thank Qingna Fan with Intel (Dalian) Ltd. for the helpful suggestions about the process of tracking issues. We thank Weiqin Zou, Jingxuan Zhang, and Jingyuan Cai with Dalian University of Technology for their help on manual validation of tag bugs. This work is partially supported by the National Natural Science Foundation of China under grants 61175062, 60805024, and 61033012, and the Fundamental Research Funds for the Central Universities under Grant No. DUT12JR02.